# Dielectric anomaly at the orbital order-disorder transition in $LaMnO_{3+\delta}$

Parthasarathi Mondal, Dipten Bhattacharya,* and Pranab Choudhury
*Electroceramics Division, Central Glass and Ceramic Research Institute, Calcutta 700 032, India*

**Abstract.**

We report a novel dielectric anomaly around the Jahn-Teller orbital order-disorder transition temperature $T_{JT}$ in $LaMnO_{3+\delta}$. The transition has been characterized by resistivity ($\rho$) vs. temperature ($T$), calorimetry, and temperature dependent X-ray diffraction studies. Measurements of complex dielectric permittivity $\varepsilon^*$ (= $\varepsilon' + i\varepsilon''$) over a low frequency range (1 Hz -10 MHz) across $T_{JT}$ reveal distinct anomaly. This observation and the reported relatively high static dielectric constant at $T$ = 0 ($\varepsilon_0$ ~ 18-20), possibly, indicate that the orbital order gives rise to intrinsic polarization that undergoes transition at $T_{JT}$. The frequency dispersion of the dielectric response at any given temperature, however, reveals that the dielectric response consists of Maxwell-Wagner component, due to interfaces, within such low frequency range. The $T_{JT}$ and the nature of the anomaly in $\varepsilon'(\omega,T)$, $\varepsilon''(\omega,T)$ at $T_{JT}$, of course, vary – from sharp upward feature to a smeared plateau and then a downward trend – depending on the $Mn^{4+}$ concentration of the sample. The observation of intrinsic dielectric response due to long-range orbital order in $LaMnO_3$ – where no ferroelectric order is possible due to the absence of off-center distortion in $MnO_6$ octahedra – may throw a new light on these classes of materials vis-à-vis multiferroic materials.

*electronic address: dipten@cgcri.res.in

## 1. Introduction

Recent progress in research on different orbital phases and their correlation with charge and spin phases sheds light on the close interplay between charge, spin, orbital, and lattice degrees of freedom in rare-earth perovskite manganites [1]. It has been realized, though not quantitatively established, that the orbital order governs the nature of the charge and spin order. The onset of orbital order at a higher temperature appears to be a necessary prerequisite for the onset of charge or spin order at a lower temperature in many cases. The temperature gap varies and even collapses for some strongly correlated systems like $LaTiO_3$ [2]. Hence, proper understanding and characterization of the orbital phases and phase transitions assumes significance. It has already been pointed out and experimentally observed that the orbital order could either be long range or short range [3]. In other words, orbital domain size may vary depending on the structural distortion. The estimation of spatio-temporal fluctuation of the orbital domains as a function of lattice distortion and temperature seems to be quite important in characterizing the orbital phases. Related to the orbital domain size and temporal fluctuation is the nature of the order-disorder transition at a characteristic transition point. We have shown recently that with the increase in geometric lattice distortion or tilt order due to decrease in average A-site radius $\langle r_A \rangle$, the latent heat associated with the transition decreases and finally reaches zero [4]. We have also noticed that irrespective of the A-site rare-earth ion, the latent heat becomes zero at a critical $\langle r_A \rangle_c \approx 1.180$ Å [5]. The pattern of the drop, however, varies with ion-type, which interestingly, can be made to collapse onto a universal pattern for a suitable choice of scaling parameter.

The broadband dielectric spectroscopy helps in characterizing the phase transitions in doped Mott insulator systems like cuprates (the parent compounds of high-$T_c$ superconductors) [6], manganites (both hole and electron doped) [7-10], other non-oxide spinel systems [11] etc, which are relatively low resistive and are not traditionally used as dielectrics. In systems where one might observe glass transition, dielectric spectroscopy assumes significance in evaluating the relaxation time scale across the phase transition point and correlating the relaxation dynamics with microscopic features. Since, in pure manganites the latent heat drops with the drop in $<r_A>$ and becomes zero [4,5], it seems that the orbital domain size also decreases with $<r_A>$ and eventually an *orbital glass* phase evolves beyond $<r_A>_c$. One can possibly distinguish between glassy dynamics and dynamics of the systems with long range order using the dielectric spectroscopy [11]. The static long-range orbital order phase should relax slowly and, therefore, can be probed by low-frequency signal while a disordered phase with short-range order is expected to relax faster and hence should be probed by higher frequency signal. The relaxation time scale may vary depending on the orbital domain size. The extension of energy scale of the probe from low to optical range helps in covering the entire spatio-temporal range of the orbital order: from long range slower modes of the entire orbital order network to faster local phonon and local orbiton modes around the Jahn-Teller active $Mn^{3+}$ ions in the optical range. The IR- and Raman-active phonon and orbiton modes have been studied in $LaMnO_3$ as a function of temperature by other authors [12-14]. However, to the best of our knowledge, no report so far exists in published literature on low-frequency dielectric response of $LaMnO_3$ near orbital order-disorder transition. Therefore, we attempt to study the phase transition features using low-frequency dielectric spectroscopy in pure $LaMnO_{3+\delta}$ with long-range orbital order as a first step towards unveiling the phase transition dynamics across the entire range of $<r_A>$.

In this paper, we report our results on dielectric anomaly at the orbital order-disorder transition temperature ($T_{JT}$) in pure $LaMnO_{3+\delta}$ having varying $Mn^{4+}$ concentration. The $T_{JT}$ as well as the nature of the anomaly varies with the $Mn^{4+}$ concentration. We probe primarily the dielectric response arising out of space charge and intrinsic polarization within the frequency window used in this work. In polycrystalline samples, interface polarization results from inhomogeneous regions due to insulating grain boundaries and conducting grains. We notice that at a given temperature the overall dielectric response, across the frequency range, is dominated, essentially, by Maxwell-Wagner relaxation dynamics arising as a result of such interfaces even though the intrinsic polarization, due possibly to orbital order, also has significant influence. The sharp feature in dielectric response at $T_{JT}$ becomes smeared with the increase in $Mn^{4+}$ concentration.

## 2. Experiments and Results

The experiments have been carried out on high quality bulk polycrystalline $LaMnO_{3+\delta}$ samples. Details of the sample preparation are given in our earlier papers [4,15]. The $Mn^{4+}$ concentration has been varied by treating the samples under different atmospheres: air, air + $N_2$ or air + Ar, and pure Ar. The orbital order-disorder transition was characterized by four-probe dc resistivity vs. temperature measurements, calorimetry using DTA-50 (Shimadzu) and Diamond DSC (Perkin-Elmer), and temperature dependent X-ray diffraction (XRD) study. The $Mn^{4+}$ concentration was estimated from chemical analysis (redox titration). The XRD patterns of the samples at different temperatures were recorded on Philips X-Pert Pro diffractometer, with Anton Parr high temperature attachment. Rectangular bar shaped samples of typical dimensions 15mm×5mm×1.5mm have been used for resistivity measurements. The current leads and voltage probes have been made of high quality platinum paste and platinum wires. The contacts were cured overnight at around 1100°C under appropriate atmosphere. The dielectric measurements have been carried out using Solartron (Model 1260) Frequency Response Analyzer (FRA) coupled with a dielectric interface (Model 1296A) across a temperature regime 300-950 K with the help of Impedance Spectroscopy software. The applied field amplitude was varied over 0.1-1 volt. However, no dependence of the dielectric response pattern could be noticed as a result of such variation. We have also studied the dielectric response at ~77 K in order to compare our data with similar systems' data measured by others [7] at ~77 K. The measurement across $T_N$ (~140 K) in pure $LaMnO_3$ reveals significant dielectric anomaly near $T_N$ that possibly reflects the spin-orbital coupling [16]. The electrodes have been made of high quality platinum, gold, and vapor deposited silver. Data obtained for different electrodes were compared in order to extract the intrinsic features of the samples. The instrument, cable, and lead contribution to the dielectric data have been carefully subtracted. Since the samples are basically low resistive, we employed constant current mode for measurement of the dielectric response [7]. However, we have noticed that interface polarization and, hence, Maxwell-Wagner contribution plays a major role in determining the dielectric permittivity and its frequency dependence in all the cases. The intrinsic capacitive response is small at room temperature. In spite of such poor response, we noticed distinct anomaly at $T_{JT}$ in both the real and imaginary part of the dielectric permittivity. Therefore, it appears that the dielectric spectroscopy can be successfully applied for studying the orbital order-disorder transition in different rare-earth perovskite manganites.

The $Mn^{4+}$ concentrations for the four samples – sample 1 ($T_{JT}$ ~ 730 K), sample 2 ($T_{JT}$ ~ 718 K), sample 3 ($T_{JT}$ ~703 K), and sample 4 ($T_{JT}$ ~653 K) – are 2%±0.3%, 5%±0.6%, 9%±0.7%, and 15%±1.2%, respectively. In Fig. 1, we show the X-ray diffraction (XRD) patterns of a representative sample (sample 1) collected at different temperatures – at room temperature (i.e., below $T_{JT}$) and above $T_{JT}$ – and their Rietveld refinement by Fullprof (ver 2.3, 1999). The refinement parameters including the results like atomic positions etc. are mentioned in Table-I. In Fig. 2, we show the variation of lattice parameters and lattice volume with $Mn^{4+}$ concentration at room temperature [17]. And the variation of the relevant parameters with temperature for the samples 1 and 4 is shown in Fig. 3. Across the transition point it is found that the low temperature orthorhombic O' phase ($c/\sqrt{2}<a<b$) becomes orthorhombic O phase ($c/\sqrt{2}>a\approx b$) at higher temperature. The temperature-dependent X-ray diffraction study yields the volume collapse ($\Delta V/V$) at $T_{JT}$ to be ~0.3% for sample 1 and ~0.016% for sample 4 [18]. The orthorhombic distortion is found to be dropping from ~2.1% to ~0.045% over a temperature regime of 300-773 K for sample 1 and from ~1.05% to ~0.13% over a temperature regime of 300-723 K for sample 4. In Fig. 4, we show the resistivity ($\rho$) vs. temperature ($T$) patterns for all the samples. The temperature range, marked by two transition points $T^*$ and $T_{JT}$, can be considered as the orbital order-disorder transition zone. $T^*$ and $T_{JT}$ define the onset [19] and completion of the transition, respectively. They are plotted as a function of $Mn^{4+}$ concentration in the top inset of Fig. 4. As expected, they drop systematically with the increase in $Mn^{4+}$ concentration. The bottom inset shows the $\ln(\rho/T)$ vs. $1/T$ plots. Linear $\ln(\rho/T)$ vs. $1/T$ pattern highlights validity of adiabatic small polaron hopping transport mechanism – $\rho(T) = \rho_0 T.\exp(E_a/k_BT)$ – over a certain temperature range. Below $T^*$ and above $T_{JT}$, the $\ln(\rho/T)$ vs. $1/T$ patterns follow straight lines. As shown in the bottom inset of Fig.4, $T^*$ marks the onset of deviation of $\ln(\rho/T)$ vs. $1/T$ pattern from straight line while $T_{JT}$ is evaluated as the point above transition at which the pattern starts following the straight line. The transition zone becomes broader with the increase in $Mn^{4+}$ concentration. $T_{JT}$ evaluated from thermal studies (differential thermal analysis or differential scanning calorimetry) nearly coincides with the $T_{JT}$ evaluated from resistivity plots thus. The uncertainty in $T_{JT}$, estimated from both resistivity and thermal studies, and in $T^*$, estimated from resistivity patterns, is mentioned in Fig. 4 top inset as error bars. The adiabatic small polaron hopping mechanism appears to be valid at below $T^*$ and above $T_{JT}$, with different activation energies. The activation energies ($E_a$), below $T^*$ and above $T_{JT}$, are estimated to be ~2550 K, ~4100 K; ~2920 K, 3180 K; ~2920 K, ~3230 K; and ~2690 K, ~3170 K for the samples with $T_{JT}$ ~ 730 K (sample 1), ~ 718 K (sample 2), ~703 K (sample 3), and ~653 K (sample 4) respectively. In Fig. 5, we show the DTA patterns depicting endothermic peaks at corresponding $T_{JT}$. While the thermograms corresponding to the samples 1, 2, and 3 are shown in the main frame, thermogram for sample 4 is shown in the top inset since the peak near the transition is almost an order of magnitude smaller. The DTA thermograms have been measured with α-alumina reference. A drift in baseline is observed in some cases because of difference in heat capacity. However, through proper baseline correction one can remove the drift. In fact, in order to display the peak near $T_{JT}$ clearly for sample 4, we refined the thermogram corresponding to sample 4 (top inset) by baseline correction. The latent heat is calculated, for all the cases, from the area under the peak after correcting the baseline of raw thermograms. In the bottom inset, we show the latent heat vs. $Mn^{4+}$ concentration. The latent heat drops with the increase in $Mn^{4+}$ concentration. However, one interesting point worth noting is that even in a sample with large $Mn^{4+}$ concentration (sample 4), the latent heat does not become zero. It has been pointed out earlier [20] that the first-order transition gives way to second-order transition for a very low $Mn^{4+}$ concentration (~2.5%) in Ba-doped $LaMnO_3$. It seems that $Mn^{4+}$ concentration is not the only factor that governs the nature of the transition; Ba-doping and consequent disorder too plays an important role in determining the nature of the transition.

In Figs. 6(a)-(f), we show the real and imaginary parts of the permittivity $\varepsilon'(\omega,T)$, $\varepsilon''(\omega,T)$ as a function of temperature for all the samples. Both in $\varepsilon'(\omega,T)$ and $\varepsilon''(\omega,T)$, a distinct anomaly could be observed at $T_{JT}$ in the case of all four samples. This observation and the reported [8] reasonably large static dielectric constant of $LaMnO_3$ at $T = 0$ can be taken as signatures of formation of orbital order induced intrinsic polarization. In this paper, we present only this relative variation of the nature of the anomaly at $T_{JT}$ by comparing the data observed in different samples.

## 3. Discussion

The pure $LaMnO_3$ contains C-type orbital-ordered phase while with the increase in $Mn^{4+}$ concentration, the volume fraction of orbital-disordered phase increases. The transition at $T_{JT}$, therefore, evolves as a function of volume fraction of the disordered phase. The transition zone bound by the temperatures $T^*$ and $T_{JT}$ also extends with the increase in $Mn^{4+}$ concentration (top inset of Fig. 4). The activation energy $E_a$ – obtained from the dc resistivity vs. temperature data by using small polaron hopping transport model – at below $T^*$ ($E_{aT^*}$) and above $T_{JT}$ ($E_{aT_{JT}}$) appears to exhibit mutually opposite trend with variation in $Mn^{4+}$ concentration: while $E_{aT^*}$ increases with the increase in $Mn^{4+}$ concentration, $E_{aT_{JT}}$ decreases. This shows that even though the charge carrier concentration increases and, therefore, the overall electrical resistivity decreases, the mobility of the charge carriers, in fact, decreases below $T^*$ with the increase in $Mn^{4+}$ concentration. On the other hand, drop in $E_{aT_{JT}}$ with increase in $Mn^{4+}$ concentration signifies higher mobility and, therefore, relevance of vibronic charge carriers (mobile vibrons formed by charge carriers coupled with locally cooperative lattice vibration) at above $T_{JT}$ [21]. The calorimetric data show how the latent heat associated with the order-disorder transition drops with the increase in $Mn^{4+}$ concentration (Fig. 5). The observation of first order transition with finite latent heat for all the samples is corroborated by the temperature dependent XRD study. In the case of sample 1, the finite lattice volume contraction (~0.3%) is observed between the points around $T_{JT}$ (Fig. 3). The contraction signifies a first order transition. In the case of sample 4, of course, a much smaller contraction (~0.016%) is observed between the points near $T^*$. In this case, the orthorhombic distortion too, decreases between room temperature and 543 K. Beyond $T_{JT}$ (~653 K) the lattice volume depicts an increase whereas the orthorhombic distortion remains small (Fig. 3).

We now concentrate on the dielectric response of the samples. Following facts are evident in the data presented in Fig. 6. (a) There are clear anomalies near the corresponding $T_{JT}$s for all the samples. (b) The nature of the anomaly at $T_{JT}$ varies with the increase in $Mn^{4+}$ concentration following a certain trend – from sharp upward feature to a smeared plateau and then a downward feature to finally, a rather broader downward peak. (c) There is a large frequency dispersion of the real and imaginary permittivity, at any given temperature, which reveals dominance of Maxwell-Wagner relaxation behavior within the frequency domain used here [22]. This is due to interfaces formed by conducting grains and insulating grain boundaries. No characteristic Debye relaxation peak could be observed in $\varepsilon''(\omega)$. (d) The pattern of temperature dependence of the real and imaginary permittivity varies from sample to sample as contribution from the intrinsic polarization varies with temperature. (e) The temperature dependence of permittivity as well as the nature of the dielectric anomaly at respective $T_{JT}$ does not vary much with variation in electrode – Pt, Au, Ag. *These are the central results of this paper.* In Fig. 6(d), we show $\varepsilon'(\omega,T)$ for sample-2 with Au electrode. The general features like rise in $\varepsilon'(\omega,T)$ with temperature and a bend and, finally, a rise beyond $T_{JT}$ are present in both the set of data – recorded for sample-2 with Pt and Au electrodes. The slight difference noticed in the feature near the anomaly could be due to difference in adhesion of the electrodes. The Au-electrode appears to offer better adhesion. In the case of sample 4, prominent anomaly near $T_{JT}$ is observed at higher frequency (≥100 kHz). The lower frequency data (not shown) reveal a kink-like anomaly and not a prominent one. This could be because at higher $Mn^{4+}$ concentration, the volume fraction of the orbital ordered phase decreases and hence, higher frequency ac probe is required in order to track the orbital order-disorder transition. In fact, the local dynamic Jahn-Teller distortion can only be studied by optical probe [23].

We first address the issue of dielectric contribution from contacts and other interfaces like grain boundaries and intrinsic interfaces, if any. The $\varepsilon'(\omega)$, $\varepsilon''(\omega)$ show large frequency dispersion within the frequency window used in this work. Both $\varepsilon'(\omega)$, $\varepsilon''(\omega)$ tend to higher values as $\omega \rightarrow 0$ without showing any clear relaxation feature. This could be due to combination of following factors: (i) large frequency dispersion of the interface polarization, (ii) non-trivial frequency dependence of intrinsic resistance (R) and capacitance (C) elements due to essentially non-Debye type dielectric relaxation, (iii) relaxor-type behavior of the intrinsic polarization. The contribution of the interface polarization normally dominates the overall response in such low-resistive polycrystalline samples. However, the complex plane analysis of the impedance spectra does not help in separating the relaxation spectrum of interface and intrinsic polarization within the temperature range explored in this work. Consequently, we could not estimate the *intrinsic* dielectric permittivity and the relaxation time scale ($\tau$) across the temperature range of this study. Even, comparison of different formalisms like, impedance ($Z'$, $Z''$), permittivity ($\varepsilon'$, $\varepsilon''$), and modulus ($M'$, $M''$) also does not help in determining these parameters. Extension of the frequency range beyond 1 MHz may help in observing full relaxation spectrum corresponding to the intrinsic polarization in the frequency domain. It is to be noted that the full relaxation spectrum corresponding to intrinsic polarization could be observed only at lower temperature (77-250 K) within the frequency window used (10 Hz-1MHz) [16]. Near 250 K, the relaxation feature shifts to even higher frequency. It is also noteworthy, in this context, that because of such dominance of Maxwell-Wagner effect, the real and imaginary permittivities assume very high value ($>10^7$). Such colossal values could be observed in other such strongly correlated systems having mobile charge carriers [6-11]. It seems that in all the cases such large values result from interfaces and, therefore, are not intrinsic. In fact, further work seems necessary in order to extract the intrinsic dielectric permittivity as well as its relaxation in these systems. Here, of course, we are concerned more in relative comparison of the data as a function of $Mn^{4+}$ concentration, in a series of pure $LaMnO_3$ systems, than the absolute values of the intrinsic permittivity. However, in order to get a fairly correct idea about the relative contribution of contacts and other interfaces to the overall dielectric response and its frequency dispersion, we carried out measurements on single crystal of $LaMnO_3$, bulk polycrystalline sample coated with alumina, and bulk samples coated with different electrodes like Au, Ag, and Pt without alumina layer. In the cases of single crystal and alumina-coated sample, Pt electrodes have been used. The single crystal is expected to be free of grain boundaries while the alumina coating on bulk sample is expected to suppress the spurious capacitance due to contacts [6]. The suppression of spurious capacitance at the contacts, of course, cannot be total as depending on the thickness ratio between the bulk and coating, one can observe finite frequency dispersion of dielectric permittivity even in this case. The dielectric spectra at room temperature for these cases are shown in Fig. 7 as an example. It is quite clear from the spectra in Fig. 7 that both single crystal and alumina-coated bulk sample exhibit lesser frequency dispersion compared to bulk un-coated samples. Comparison of these data with the data obtained for bulk polycrystalline samples with different electrodes (Au, Ag, Pt) and without alumina coating provides a rough estimation of the relative contribution of the contacts and other interfaces. It appears that nearly two orders of magnitude dispersion in dielectric permittivity, over a frequency range of four orders, results from the intrinsic polarization contribution. Therefore, we conclude that not only Maxwell-Wagner relaxation due to interfaces there is a *sizable component of intrinsic polarization dispersion* as well, across the entire frequency window used here. This dispersion could result from either non-trivial frequency dependence of R and C or relaxor-like behavior or a combination of both. At any given frequency within this range, the overall dielectric permittivity is given by the summation of intrinsic and interfacial permittivity normalized by ratio of logarithm of volumes (Lichteneckar's rule).

We now turn our attention to the nature of the dielectric anomaly at $T_{JT}$ and its variation with the increase in $Mn^{4+}$ concentration. The increase in $Mn^{4+}$ concentration gives rise to a decrease in volume fraction of the orbital ordered phase that, in turn, results in such evolution in the nature of the anomaly. The shift from upward feature to a downward one is intriguing. Upward anomaly is observed in multiferroic $TbMnO_3$, $DyMnO_3$ [24] where as in the case of $BiFeO_3$, one observes downward anomaly [25] and its smearing with the increase in non-stoichiometric oxygen [26]. In all these cases, of course, the dielectric anomalies are observed near Neel point $T_N$ and signify magneto-electric coupling. In the present case, the dielectric anomaly near $T_{JT}$ signifies transition in orbital-order driven electrical polarization. The quantitative estimation of the anomaly as a function of $Mn^{4+}$ concentration is possible only if we could correlate the orbital domain size with the local electrical polarization length in samples having different $Mn^{4+}$ concentration. The orbital domain size is expected to become smaller with the increase in $Mn^{4+}$ concentration that, in turn, is expected to give rise to a systematic variation in electrical polarization length scale. This detailed study is beyond the scope of the present work.

Finally, we point out that for non-polar systems, one observes the static dielectric constant ($\varepsilon_0$) at $T = 0$ to be within 1-5. It has already been reported [8], however, that for $LaMnO_3$, $\varepsilon_0$ is within 18-20, which is reasonably high. It gives an indication that probably intrinsic polarization develops *locally*. Since $LaMnO_3$ is *globally* centrosymmetric, it cannot exhibit very high dielectric constant in global scale. However, because of cooperative Jahn-Teller effect together with tilt order due to La-site radius, one can predict *local* breakdown of the centrosymmetry and development of dipoles due to charge separation in Mn-O plane. The correlation among local Mn-O bond distortion due to orbital order and charge displacement giving rise to local dipoles has been established in Ref. 16. Another important point worthy of mentioning here is that both temperature dependence of $\varepsilon'(\omega)$ and the anomalies at $T_{JT}$ vary from sample to sample. This is unexpected in an essentially non-polar system. The volume fraction of orbital ordered phase certainly has a role to play here. Piecing together all these features, it is possible to conclude that long-range orbital gives rise to non-trivial intrinsic polarization even in a globally centrosymmetric compound like $LaMnO_3$. Yang *et al.*[27] have pointed out the relevance of orbital-order driven polarization in $BiMnO_3$ very recently. Another example is the electronic mechanism (charge order) driven polarization that is found to have given rise to long-range ferroelectric order in $LuFe_2O_4$ [28]. Our observation of the dielectric anomaly within the low frequency range (100 Hz – 1 MHz) together with large frequency dispersion and frequency dependence of the nature of the anomaly points out that intrinsic polarization develops, at least, *locally* because of orbital order. Additional structural distortion due to either stereo-chemical activity of Bi $6s^2$ lone pair in $BiMnO_3$ and $BiFeO_3$ or $MnO_5$ bi-pyramid in $RMnO_3$ (R = Dy, Tb, Y, Yb, Lu etc.) or electron correlation effect in charge-ordered $LuFe_2O_4$ gives rise to global ferroelectricity. The long-range polarization length scale may vary depending on the presence or absence of the additional mechanism along with the long-range orbital order. Alongside, the extent of dispersion in intrinsic dielectric permittivity also varies depending on the polarization length scale. It will be interesting to study how the polarization length scale grows as a result of systematic increase in structural distortion in presence of long-range orbital order. This will be attempted in a future study.

**4. Summary**

In summary, we have indicated that there is a possibility of observing orbital order driven intrinsic electrical polarization in pure $LaMnO_{3+\delta}$. As a result, one observes distinct dielectric anomaly at the transition temperature $T_{JT}$. The Maxwell-Wagner effect arising out of interfaces and/or mobile charge carriers, of course, dominates the overall dielectric response in the temperature and frequency range covered in this work. However, the temperature dependence of the real and imaginary permittivity as well as the variation in the nature of the anomaly at $T_{JT}$ from sample to sample reveals influence of intrinsic orientation polarization due to orbital order. The transition temperature itself and the latent heat associated with the transition decrease with the $Mn^{4+}$ concentration yet interestingly they do not become zero, at least, over the range of $Mn^{4+}$ concentrations studied here. This shows that one can play with relative volume fraction of orbital ordered and disordered phases and learn about the phase transition from dielectric relaxation across the $T_{JT}$ in all such insulating manganites exhibiting orbital order-disorder transition. Further work at even higher frequency range (>1 MHz) seems to be necessary in order to estimate the intrinsic dielectric constant as well as its relaxation dynamics. Such results may spur more work aimed towards discovering the driving mechanism behind intrinsic polarization observed in these centrosymmetric compounds.

**Acknowledgements**

We acknowledge help of P.S. Devi of CGCRI in estimation of $Mn^{4+}$ concentration by chemical analysis and A.K. Tyagi of BARC, Mumbai, India for carrying out the X-ray diffraction studies at different temperatures. We also thank P. Mandal of SINP, Calcutta, India for providing the single crystal of $LaMnO_3$.

Table-I. Results of the Rietveld refinement of the X-ray diffraction patterns for samples 1 and 4 at few selective temperatures below and above transition ($T_{JT}$). The atomic positions are: La (x, y, 0.25), Mn (0.5, 0, 0), O1 (x, y, 0.25), O2 (x, y, z). The space group is Pbnm.

| | T (K) | La | | O1 | | O2 | | | $R_p$ | $R_{wp}$ | $\chi^2$ |
|---|---|---|---|---|---|---|---|---|---|---|---|
| | | x | y | x | y | x | y | z | | | |
| Sample 1 | 300 | 0.99334 | 0.04599 | 0.05062 | 0.50187 | 0.72115 | 0.30422 | 0.02602 | 13.8 | 17.0 | 3.49 |
| ($T_{JT}$ ~730 K) | 673 | 0.9939 | 0.04072 | 0.0683 | 0.49388 | 0.71684 | 0.2986 | 0.0282 | 20.1 | 28.7 | 1.20 |
| | 773 | 0.99377 | 0.02631 | 0.06205 | 0.50085 | 0.78503 | 0.22204 | -0.0181 | 19.4 | 28.6 | 1.15 |
| Sample 4 | 300 | 0.99487 | 0.03569 | 0.04833 | 0.49967 | 0.74036 | 0.28979 | 0.03289 | 14.9 | 20.3 | 3.22 |
| ($T_{JT}$ ~ 653 K) | 543 | 0.99331 | 0.02405 | 0.03882 | 0.50070 | 0.73557 | 0.26752 | 0.01676 | 14.4 | 18.7 | 2.55 |
| | 723 | 0.98933 | 0.01344 | 0.0189 | 0.49735 | 0.75026 | 0.30001 | -0.0002 | 17.8 | 23.6 | 3.96 |

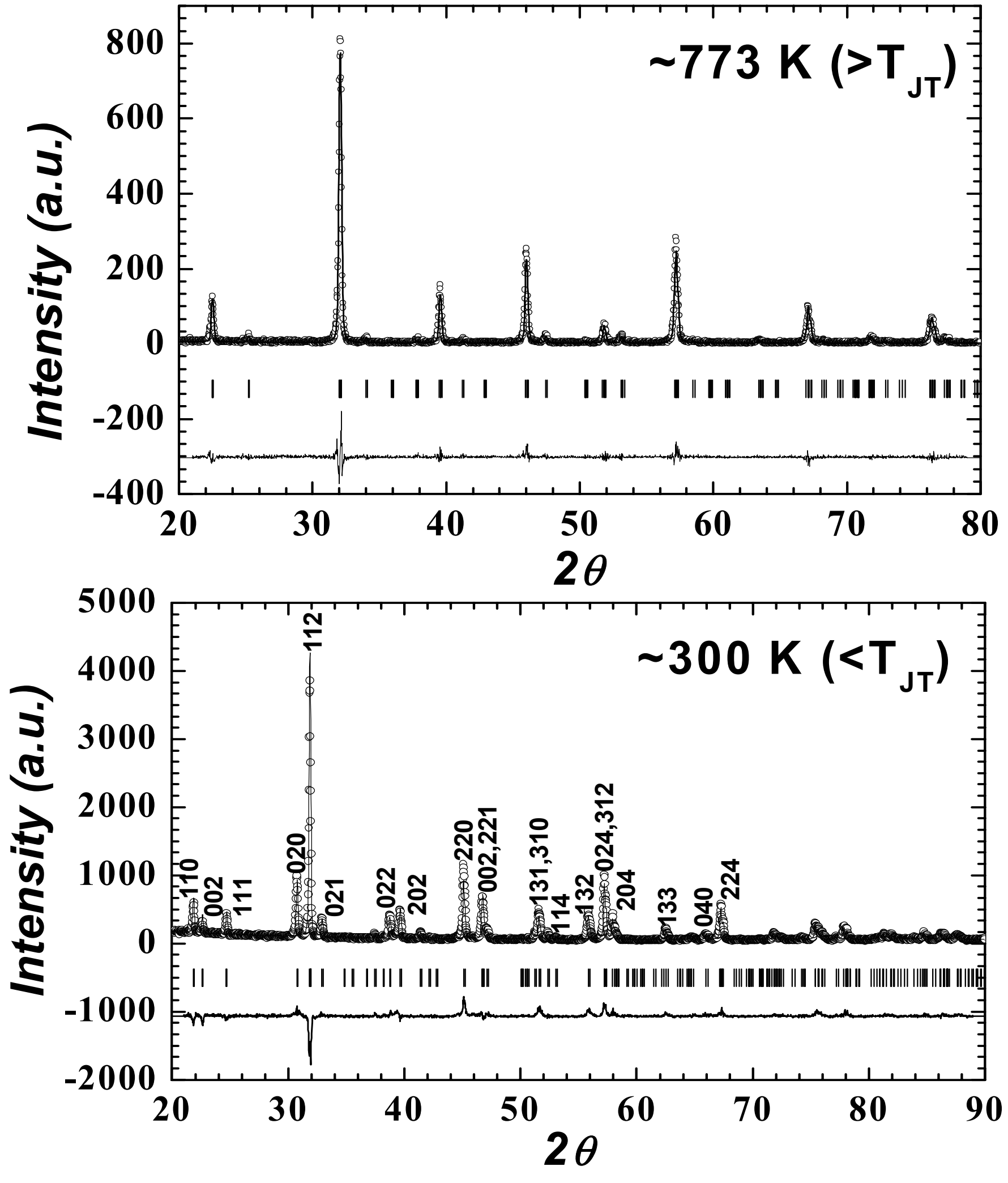

Fig. 1. X-ray diffraction patterns – refined by Fullprof (ver 2.3, 1999) – for sample 1 at below and above $T_{JT}$ (~300 and 773 K). The open circles denote the experimental data points.

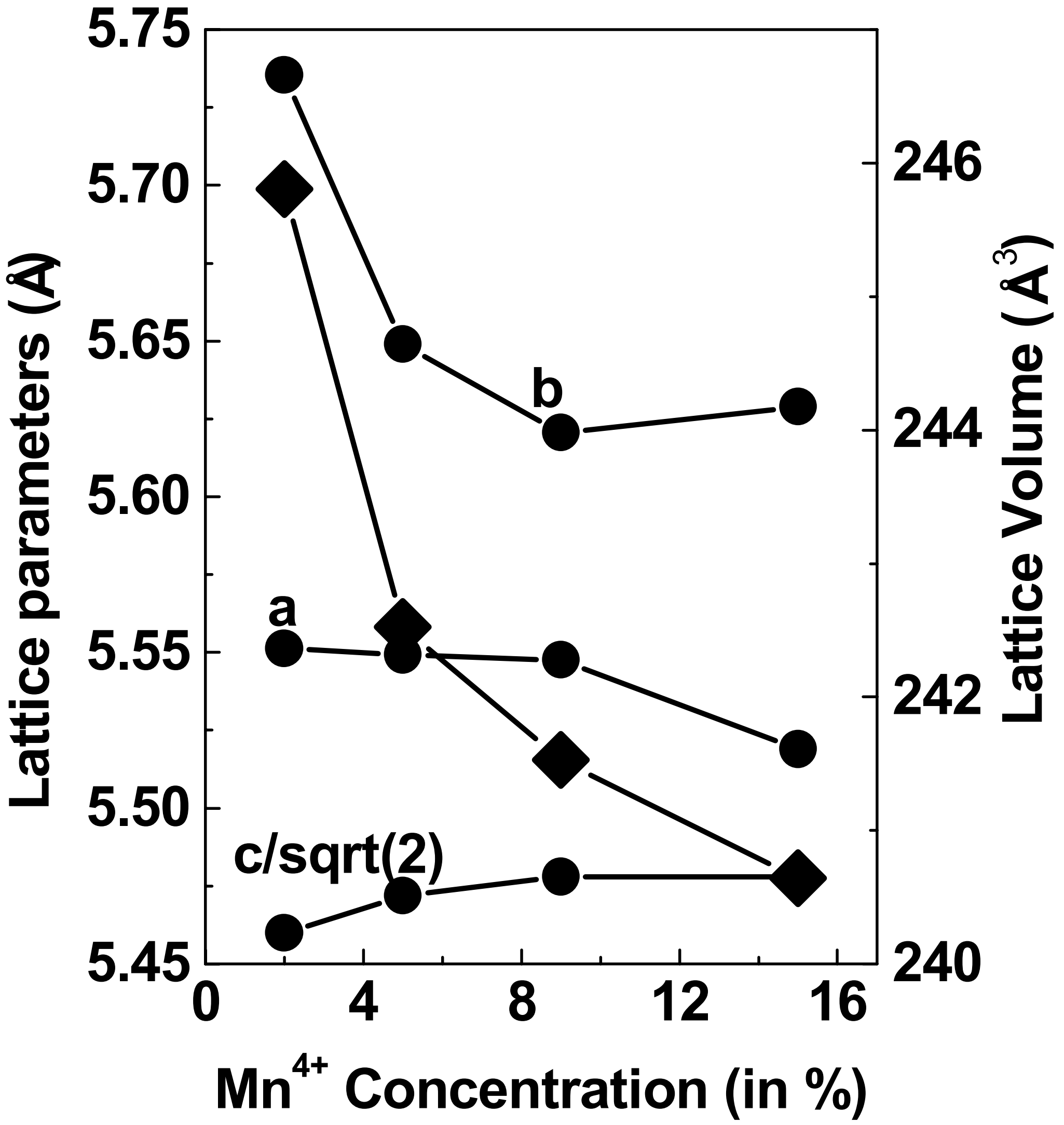

Fig. 2. Variation of lattice parameters and lattice volume (♦) at room temperature as a function of $Mn^{4+}$ concentration in $LaMnO_{3+\delta}$.

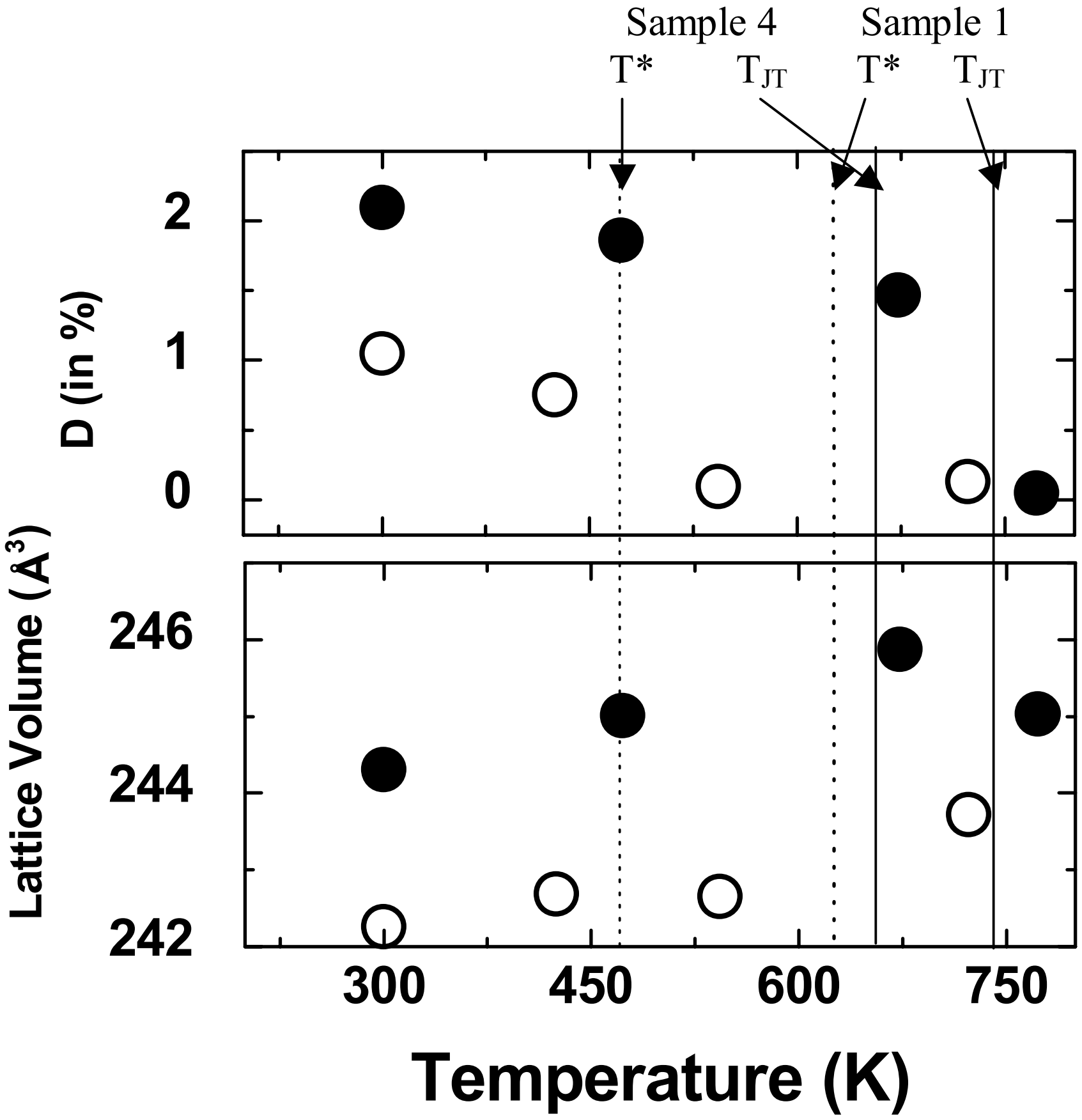

Fig. 3. Variation of lattice volume and orthorhombic distortion D as a function of temperature for samples 1 (●) and 4 (○). The orthorhombic distortion D is given by $D = [|(a - a_l)| + |(b - a_l)| + |(c/\surd 2 - a_l)|]/3a$, where $a_l = (abc/\surd 2)^{1/3}$. The transition temperatures $T^*$ and $T_{JT}$ are marked as identified from resistivity vs. temperature data.

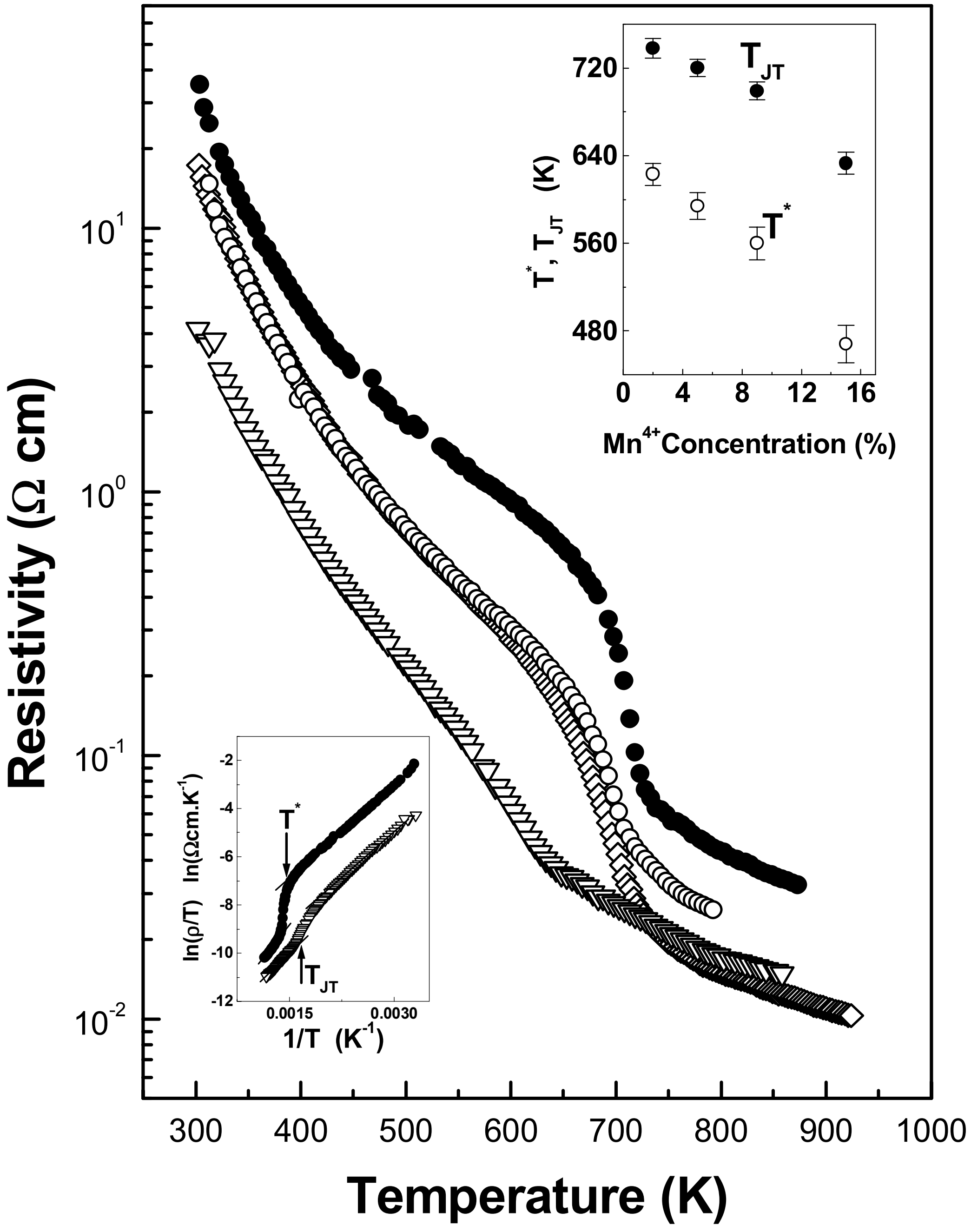

Fig. 4. Resistivity vs. Temperature plots for sample 1 (•; $T_{JT}$~730 K); sample 2 (○; $T_{JT}$~718 K); sample 3 (◊, $T_{JT}$~700 K); sample 4 (∇, $T_{JT}$~653 K). Top inset shows the variation in $T^*$, $T_{JT}$ with $Mn^{4+}$ concentration and bottom inset shows ln(ρ/T) vs. 1/T plot for samples 1 and 4.

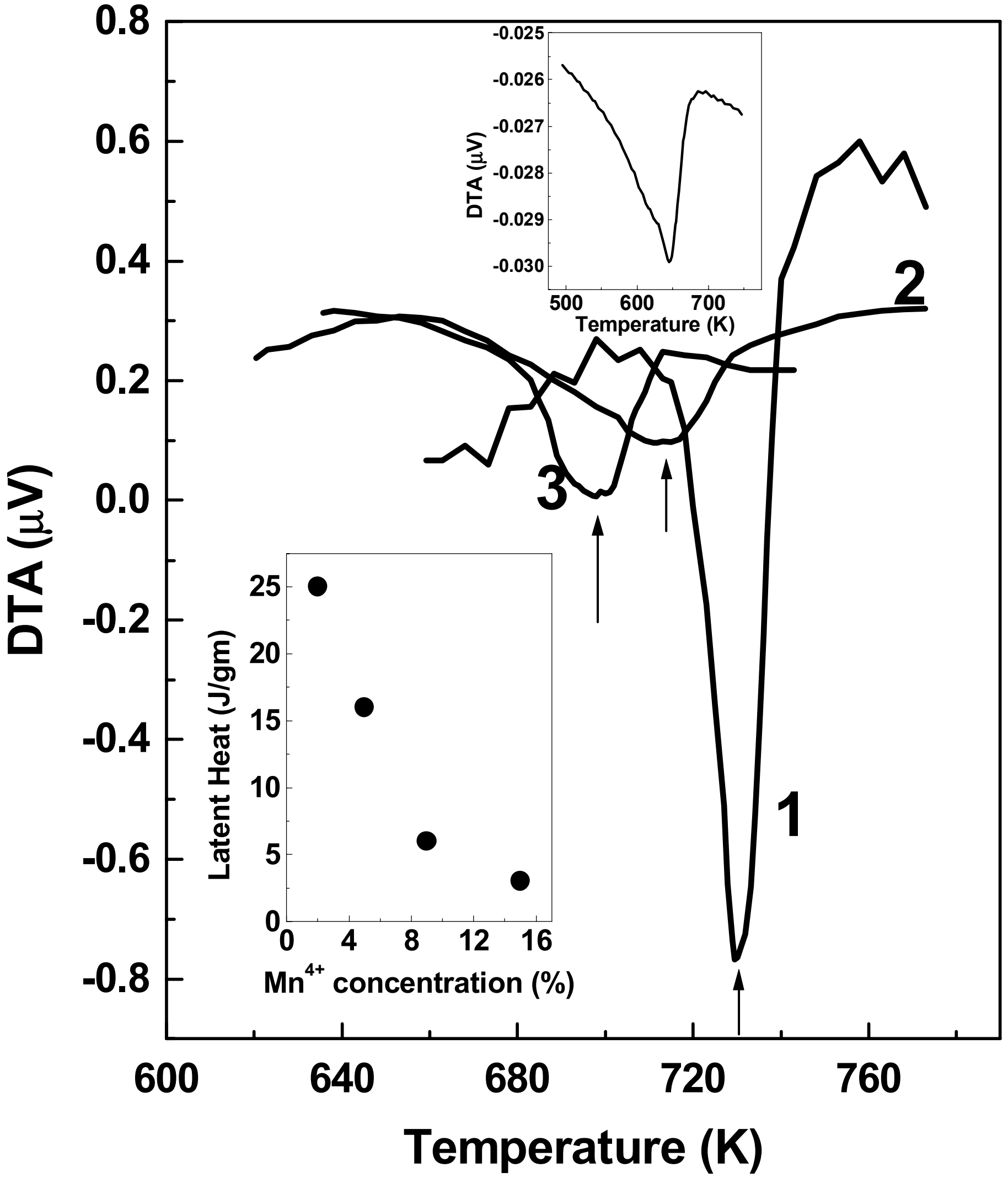

Fig. 5. DTA thermograms for the samples 1, 2, and 3. Top inset shows the thermogram for samples 4 and bottom inset shows the variation in latent heat with $Mn^{4+}$ concentration.

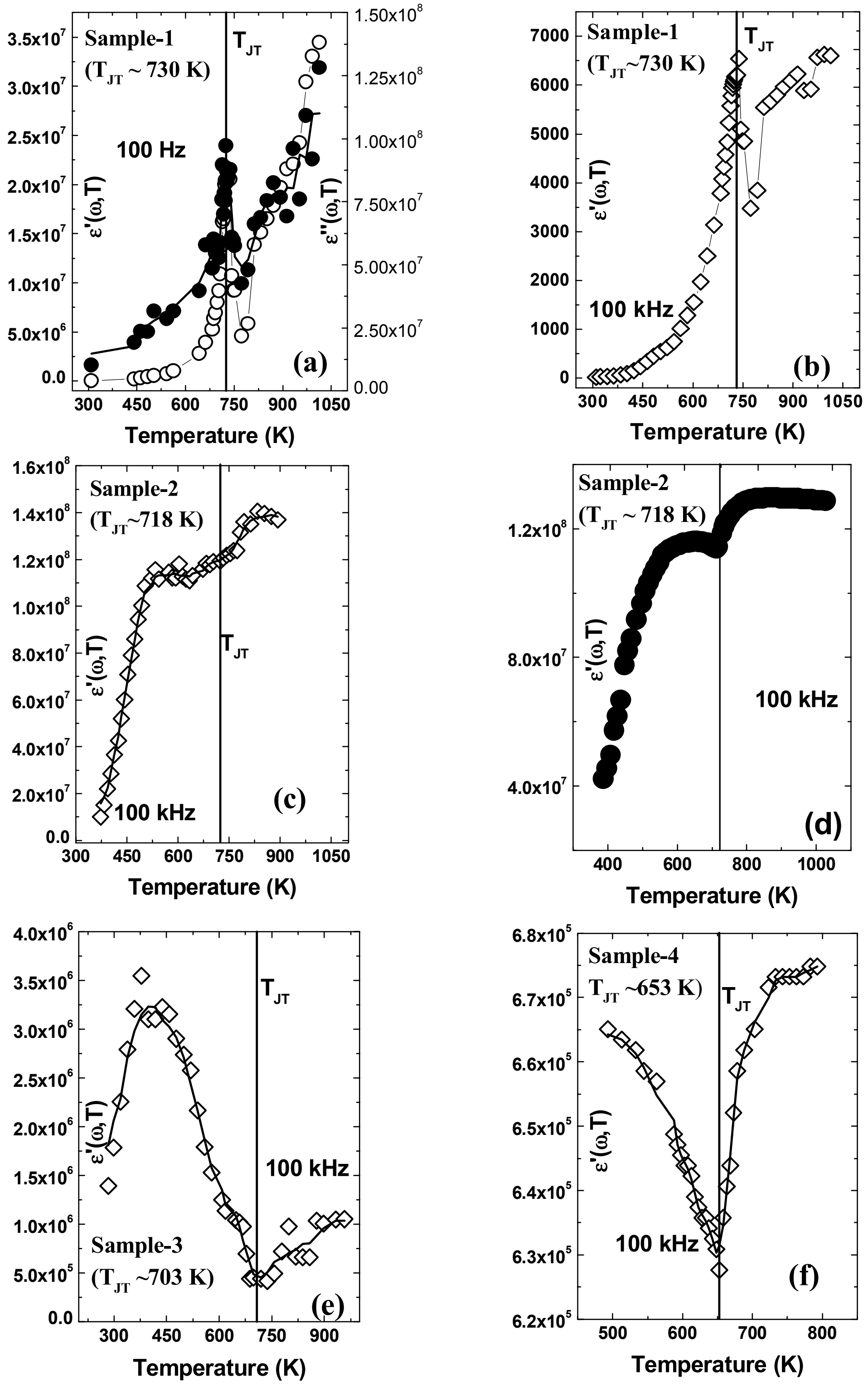

Fig. 6. Real part of the dielectric permittivity for samples 1 to 4. In frame (a) the imaginary part of the dielectric permittivity, for sample 1, is also shown (solid symbols) for example. In frame (d) the data from Au-coated sample-2 is shown. Data corresponding to other frames have been obtained from samples with Pt electrodes.

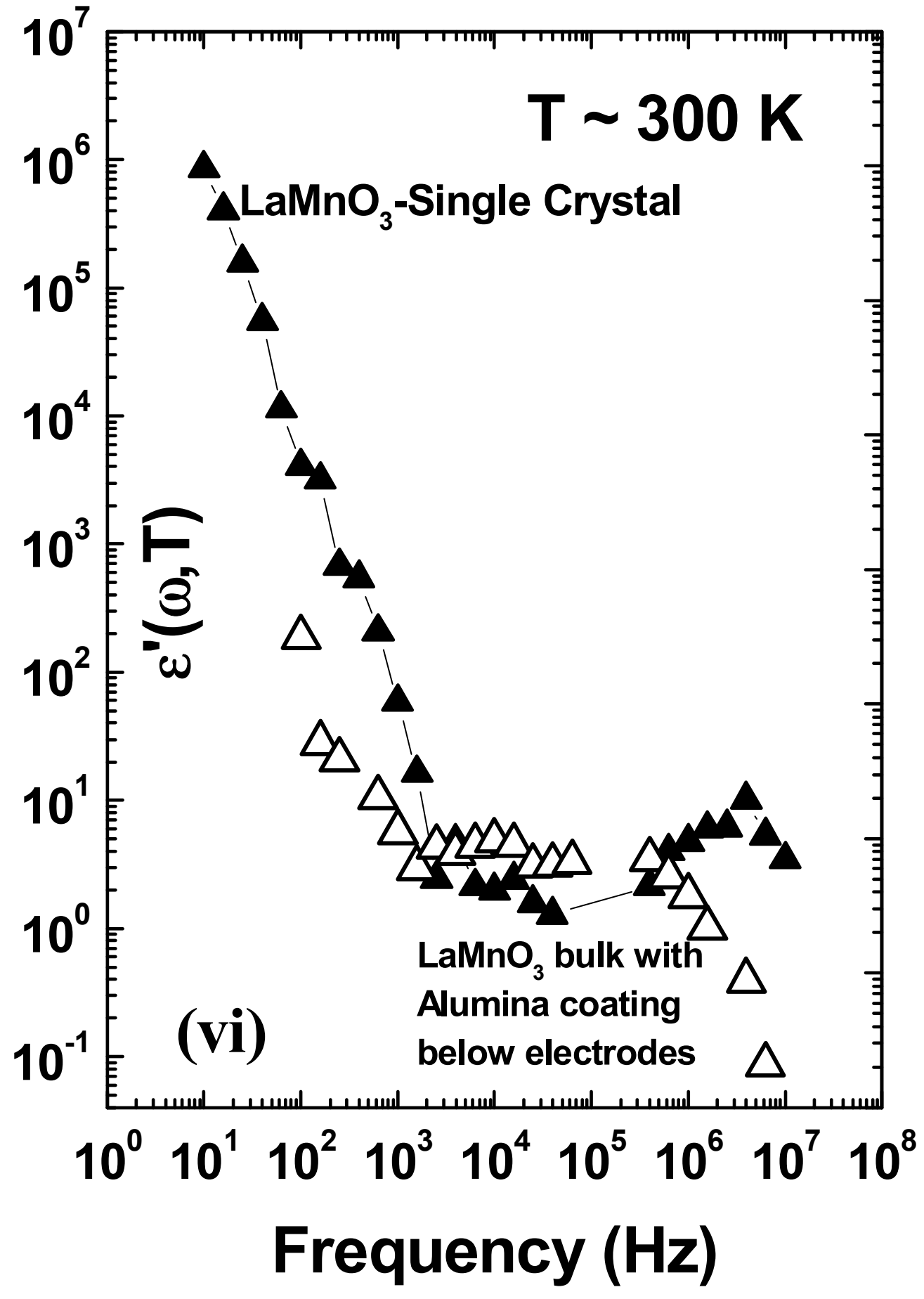

Fig. 7. The typical frequency dispersion pattern of real part of the dielectric permittivity observed in single crystal $LaMnO_3$ and bulk polycrystalline sample coated with alumina. In both the cases Pt electrode has been used.